\documentstyle[preprint,aps,epsf]{revtex}
\tighten

\def\H{{\rm H}}
\def\I{{\rm I}}
\def\inf{{\rm inf}}
\def\max{{\rm max}}
\def\min{{\rm min}}
\def\P{{\rm P}}
\def\rh{{\rm reh}}
\def\rmd{{\rm d}}
\def\rme{{\rm e}}

\def\therm{{\rm therm}}

\begin{document}
\preprint{\small OUTP-00-41P (rev)}
\draft
\title{Low-scale inflation}
\author{Gabriel Germ\'an}
\address{Centro de Ciencias F\'{\i}sicas, Universidad Nacional 
Aut\'onoma de M\'exico, Apartado Postal 48-3, 62251
Cuernavaca, Morelos, M\'exico}
\author{Graham Ross and Subir Sarkar}
\address{Theoretical Physics, University of Oxford,
1 Keble Road, Oxford OX1 3NP, UK}
\bigskip
\date{\today}
\maketitle

\begin{abstract}
We show that the scale of the inflationary potential may be the
electroweak scale or even lower, while still generating an acceptable
spectrum of primordial density perturbations. Thermal effects readily
lead to the initial conditions necessary for low scale inflation to
occur, and even the moduli problem can be evaded if there is such an
inflationary period. We discuss how low scale inflationary models may
arise in supersymmetric theories or in theories with large new space
dimensions.
\end{abstract}
\bigskip

\widetext

\section{Introduction}

Inflation provides an attractive context for discussion of the initial
conditions of the hot big bang cosmology, as well as a very plausible
mechanism for generation of the scalar density perturbations which
have left their imprint in the anisotropy of the cosmic microwave
background (CMB) and grown into the observed large-scale structure
(LSS) of galaxies \cite{books,review}. Field theoretical models of
inflation are typically of the `slow-roll' type in which an inflaton
field, $\phi^*$, evolves along a quasi-flat potential to its global
minimum. During the roll the vacuum energy is approximately constant
and drives an era of exponential growth in the cosmological scale
factor $a$. However such field theory models must explain why the
potential is extremely flat even in the presence of radiative
corrections --- the ``$\eta$-problem'' --- and also explain why the
inflaton field initially lies far from its true minimum.

Quite generally, we may expand the inflaton potential about the value
of the field $\phi^*_\I$ just before the start of the observable
inflation era when the scalar density perturbation on the scale of our
present Hubble radius \footnote{Numerically this is
$H_0^{-1}\simeq3000h^{-1}$~Mpc, where
$h\equiv\,H_0/100$\,km\,s$^{-1}$\,Mpc$^{-1}\sim0.65\pm0.15$ is the
present Hubble parameter. LSS and CMB observations can measure the
density perturbation down to galactic scales ($\sim1$~Mpc), a spatial
range corresponding to just 7--8 e-folds of inflation.} was generated,
and expand in the field $\phi\equiv\phi^*-\phi^*_\I$
\cite{reconstruct}. Since the potential must be very flat to drive
inflation, $\phi$ will necessarily be {\em small} while the observable
density perturbations are produced, so the Taylor expansion of the
potential will be dominated by low powers:
\begin{equation}
V(\phi) = V(0) + V^{\prime}(0)\phi + \frac{1}{2}V^{\prime\prime}(0)\phi^2 
          + \ldots
\label{expand}
\end{equation}
The first term $V(0)$ provides the near-constant vacuum energy driving
inflation while the $\phi$-dependent terms are ultimately responsible
for ending inflation, driving $\phi$ large until higher-order terms
violate the slow-roll conditions needed for inflation. These terms
also determine the nature of the density perturbations produced, in
particular the departure from an exactly scale-invariant spectrum.

The density perturbation at wavenumber $k$ is given by \cite{review}
\begin{equation}
\delta_\H^2 (k) = 
 \frac{1}{75\pi^2}
 \frac{V(\phi^*=\phi^*_\H)^3}{V^{\prime}(\phi^*=\phi^*_\H)^2M^6}\ ,  
\end{equation}
where $\phi^*_\H$ is the value of the inflaton field when the relevant
scale `exits the horizon', i.e. when $k=aH$,
$H\equiv\dot{a}/a\simeq(V/3M^2)^{1/2}$,
$M\equiv\,M_\P/\sqrt{8\pi}\simeq2.44\times10^{18}$~GeV. The slope of
the potential is given by Eq.(\ref{expand}) as
$$
V^{\prime}(\phi^*=\phi^*_\H) = V^{\prime}(0) + V^{\prime\prime}(0)\phi_\H 
                        + \ldots ,
$$
where $\phi_\H\equiv\phi^*_\H-\phi^*_\I$. If the term linear in $\phi$
in the potential (\ref{expand}) dominates when the observed density
perturbations are generated (``linear inflation''), then only the
first term above is important and the scale of the inflationary
potential is required to be near the Planck scale. This can be seen by
writing the slope as $V^{\prime}(0)=cV(0)/M$ which gives
\begin{equation}
V^{1/4}(\phi=0) \simeq (75\pi^2\delta_\H^2)^{1/4}c^{1/2}M 
                \sim 2\times10^{-2}\sqrt{c}M ,  
\label{scale}
\end{equation}
using the COBE determination of $\delta_\H\simeq2\times10^{-5}$ on the
scale of the present Hubble radius \cite{cobe}. We see that the
inflationary scale is below the Planck scale (so a field-theoretic
description may be used) but, unless $c$ is unnaturally small, will be
not too far below it \cite{books,review}.

Examples of models generating linear inflation abound. Consider for
example the commonly used `chaotic inflation'\footnote{The term
`chaotic' was in fact originally intended to refer to the initial
conditions for the inflaton, with inflation occuring either at
large \cite{chaotic} or small $\phi$ \cite{quad} field values. However
it has come to be associated exclusively with large-field models with
a generic monomial potential, which we discuss above. To avoid
confusion we specifically refer to `large-field' models where
relevant. Note that when inflation occurs at small $\phi$ the initial
conditions may alternatively be set by thermal effects
\cite{new}. This was originally found to be difficult to implement
(hence the proposal of chaotic initial conditions \cite{chaotic}), but
as we shall see these difficulties are circumvented if the
inflationary scale is low.} models in which the inflationary potential
is dominated by a monomial, $V\propto\phi^{*n}$, and $\phi^*_\I$ is
{\em larger} than the Planck scale \cite{chaotic}. The terms in the
Taylor expansion satisfy:
\begin{equation}
V^{p+1}(\phi=0)\phi^p/V^\prime(\phi=0) \simeq (\phi/\phi^*_\I)^p
                                       \ll 1.
\end{equation}
Consequently the linear term dominates the $\phi$ dependence showing
that any such large-field inflation model with a monomial potential
actually falls into the class of {\em linear} inflation models defined
above. Hence the inflationary scale is not too far below the Planck
scale, e.g. for the commonly adopted potential
$V=\case{1}{2}m^2\phi^{*2}$ one requires $m\lesssim10^{-4}M$.

There is a circumstance, however, in which it is possible in a {\em
natural} way to avoid the conclusion that the inflationary scale is
large. This is the case if the linear term is anomalously small. If,
as is the usual expectation, the field $\phi^*$ transforms under some
symmetry of the underlying theory, a linear term is forbidden if one
expands the potential about the symmetric point. In this case the
natural expectation is that the next term, namely the quadratic term
will dominate. In the large-field inflation model just discussed, this
situation does not arise because the initial value of the inflaton
field corresponds to a (chaotic) value, and any symmetry under which
it is charged is broken. However, as we shall discuss, it is possible
that the initial conditions are such as to set the inflaton field
close to the origin where the symmetry is unbroken, as in `new
inflation' \cite{new}. Provided $\phi^*_\I$ is also close to the
origin, the term linear in $\phi$ in the potential (\ref{expand}) will
indeed be very small, so the quadratic term may dominate.

It is the purpose of this paper to investigate this possibility
(``quadratic inflation'') in detail. Following the same procedure as
above we may determine the density perturbations for the case of
quadratic inflation. Writing
$V^{\prime\prime}(0)=\tilde{c}V(0)/M^2$ we now find
\begin{equation}
\delta_\H^2 \simeq \frac{V(0)^3}{75\pi^2\tilde{c}^2V(0)^2\phi_\H^2M^2}\ . 
\end{equation}
Note the appearance of the value, $\phi_\H$, of the (rescaled)
inflaton field at the time of production of the observed density
perturbations. Because of this, the scale of the inflationary
potential now depends on $\phi_\H$:
\begin{equation}
V(\phi=0)^{1/4} \simeq 2\times10^{-2}\sqrt{\tilde{c}}\phi_\H^{1/2}M^{1/2} .
\end{equation}
Thus if $\phi_\H/M$ is small, the scale of inflation will also be
small even for natural values of $\tilde{c}\sim1$. It is clear that
quadratic inflation depends sensitively on the terms in the potential
responsible for ending inflation and determining $\phi_\H$, and that
if these terms generate a small value for $\phi_\H/M$ then we have a
plausible mechanism for low-scale inflation.\footnote{A potential with
a leading quadratic term was in fact first studied in
Ref.\cite{quad}. However it was the steepening of the potential due to
this leading term that was assumed there to end inflation --- this
implies a relatively high value for $\phi_\H$, hence does {\em not}
permit a low inflationary scale.}

Apart from being of quite general interest quadratic inflation has
potential advantages for inflation in the case of models with new
large dimensions. It has been shown that such models can evade the
hierarchy problem associated with the existence of very large mass
scales because in these models the Planck scale is no longer a
fundamental quantity, instead all fundamental scales are of order the
electroweak scale. In these models, however, inflation must be
achieved via a potential which has no large scales and in this context
the quadratic inflationary potential is {\em necessary} to generate
acceptable slow-roll inflation. As we shall discuss models with new
large dimensions also offer a new way of solving the $\eta$-problem.

The alternative explanation of the hierarchy problem is that a new
symmetry --- supersymmetry --- is a good symmetry at low energy
scales. As we have remarked earlier \cite{ssi} supersymmetry provides
a very natural source of viable inflatonary potentials. Supersymmetry
prevents large radiative corrections to the potential and thus
provides a consistent framework to address the
$\eta$-problem. Furthermore, since thermal effects necessarily break
supersymmetry, there is also a very natural explanation for why the
inflaton field should have its initial high-temperature minimum far
from its zero temperature supersymmetric minimum. In quadratic
inflation models the scale of inflation can readily be identified with
the scale of supersymmetry breaking.

In this paper we study the construction and implications of quadratic
inflationary models, paying particular attention to the mechanisms for
solving the $\eta$-problem, both in the context of supersymmetric
models and in models with large new dimensions. In
Section~\ref{quadratic} we discuss the conditions that must be met by
a quadratic inflaton potential in order to generate viable
inflation. In Section~\ref{model} we introduce a simple
parameterisation of a model capable of satisfying these conditions and
study in detail in Section~\ref{characteristics} the characteristics
of quadratic inflation. In Section~\ref{sugrasection} we discuss how
the quadratic inflationary potential can arise in supergravity models
and in Section~\ref{dim} we discuss the same question in the context
of theories with large new dimensions.

\section{Quadratic Inflation}\label{quadratic}

Quadratic inflation requires that the inflaton field, $\phi$, rolls
from the origin with an inflationary potential dominated by the
quadratic term in a Taylor expansion. In field theory models such
structure is quite a natural one because scalar fields often carry
quantum numbers under a symmetry such that, in the limit where the
symmetry is unbroken, a linear term in the potential is forbidden and
the quadratic term dominates. Of course it is necessary to show that
the theory initially starts with the symmetry unbroken,
i.e. $\phi=0$. We shall demonstrate that this can happen quite
naturally through thermal effects because the dimensionless couplings
of the theory necessarily respect the symmetry and give a thermal
potential which has a minimum for vanishing $\phi$. It is also
necessary that at low temperatures the inflaton starts to roll,
spontaneously breaking the underlying symmetry. Again, as we shall
discuss, this is quite natural as the quadratic mass term often has a
negative sign, triggering spontaneous breaking of the symmetry.
Provided this mass term is small it will not affect the high
temperature potential significantly but will generate the slow-roll
inflation at late times once the temperature drops sufficiently.

Of course this explanation requires that the system initially be in
thermal equilibrium and this is not normally the case in slow-roll
inflationary models, particularly since the inflaton should be very
weakly coupled in order not to spoil the required flatness of its
potential. However quadratic inflation is special in as much as the
value of the potential during inflation is not strongly constrained by
the need to generate the correct magnitude of density
perturbations. For the case that the potential energy driving
inflation is low, we will show that the processes leading to thermal
equilibrium {\em do} have time to establish equilibrium before the
inflationary era starts.

A further crucial question is why the universe should initially be
sufficiently homogeneous for slow-roll inflation to begin. In chaotic
(large-field) inflation models \cite{chaotic}, inflation begins when the
scale of the potential energy is of order the Planck scale and the
horizon (the scale over which the universe must be homogeneous for
inflation to start \cite{homogen}) is also of order the Planck
scale. By contrast in the models discussed here, inflation starts much
later when the horizon contains many such Planck scale horizons and,
in this case, it is difficult to understand how the necessary level of
homogeneity can be realised. However this argument is not really a
criticism of the possibility that there be a late stage of (quadratic)
inflation but rather a statement that this cannot be all there
is. Thus we require that there was some other process which ensured
the necessary homogeneity at the beginning of quadratic inflation, as
in the eternal inflation scenario \cite{eternal}, or perhaps through
some quantum cosmological process \cite{qc}. Such considerations
suggest a situation in which a homogeneous universe emerges at the
Planck era and potential energy is released reheating the universe
and, as described above, setting the conditions for further periods of
inflation to occur. If the late stage of inflation generates a
sufficient number of e-folds of inflation, these earlier eras will not
have observable consequences for our universe although they will have
been crucial in setting the correct initial conditions for it to
occur.

In a complete field theory description of the fundamental interactions
there are usually many scalar fields, all candidates for generating a
period of inflation. For this reason we consider quadratic inflation
to be a generic feature. Indeed one might expect several inflationary
periods to be encountered in the evolution of the universe due to
several scalar fields slow-rolling to their minima. Again only the
last such era will be relevant to observation if it generates a
sufficiently large ($\sim50-60$) number of e-folds of
inflation.\footnote{Multiple short bursts of inflation may also be
viable \cite{us97}, as is indeed suggested by recent CMB and LSS data
\cite{us00}.}

\subsection{Slow-roll conditions}

Let us turn now to a more detailed discussion of the conditions that a
quadratic potential must satisfy if it is to generate
inflation. Starting from $\langle\phi\rangle\sim0$ and assuming that
the symmetry properties of the model forbid a linear term, the
quadratic term will dominate, giving a new inflation \cite{new}
potential of the form\footnote{In what follows we will often just use
$\phi$ to denote its vacuum expectation value (vev).}
\begin{equation}
V(\phi) \sim \Delta^4 - \frac{1}{2}m_\phi^2\phi^2 + \dots ,
\label{newinflation}
\end{equation}
where the constant vacuum energy $\Delta^4$ is now the leading term in
the potential.  

The slow-roll condition is given by \cite{review,turner}
\begin{equation}
\epsilon \equiv \frac{M^2}{2}\left(\frac{V^\prime}{V}\right)^2 \ll 1,
 \qquad |\eta| \equiv M^2\left|\frac{V^{\prime\prime}}{V}\right| \ll 1\ ,
\label{slowroll}
\end{equation}
where the potential determines the Hubble parameter during inflation
as, $H_{\inf}\equiv\dot{a}/a\simeq(V/3M^2)^{1/2}$. Inflation ends
(i.e. $\ddot{a}$, the acceleration of the cosmological scale factor,
changes sign from positive to negative) when $\epsilon$ and/or
$|\eta|$ becomes of ${\cal O}(1)$. From Eq.(\ref{slowroll}) we have
now a constraint on the mass
\begin{equation}
m_\phi \ll H_\inf \sim \Delta^2/M, 
\label{massbound}
\end{equation}
which is much smaller than its natural value --- the
$\eta$-problem. To solve this problem we are driven to consider
theories in which $m_{\phi}$ is prevented from becoming large.

\subsection{Initial conditions for inflation}
\label{initial}

What about the initial conditions necessary for inflation to commence?
We assume that some process at the Planck scale, presumably quantum
cosmological in nature \cite{qc}, creates a homogeneous patch of
space-time of sufficient size for our last stage of inflation to occur
if the inflaton field satisfies the constraints just
discussed.\footnote{In the case of additional dimensions this should
presumably occur in the underlying higher dimensional theory at the
higher dimensional Planck scale. The later stage of quadratic
inflation occurs at a scale below the compactification scale when the
effective theory is four-dimensional.}

During the Planck era the inflaton field has a natural value of ${\cal
O}(M_\P)$ due to gravitational interactions which are strong at this
scale. At later times, while $\phi$ is still large, its potential will
be dominated by the highest powers, $\phi^p/M^{\prime p-4}$, in the
inflaton potential (where $M^{\prime}$ is a high mass scale in the
theory). Such terms will cause the inflaton to flow towards the
origin. However they cannot drive $\phi$ to a sufficiently small value
for quadratic inflation to occur because they rapidly become
negligible as $\phi$ becomes small. For this reason it is necessary to
consider whether there exists some other mechanism capable of driving
$\phi$ small. In particular the thermal potential following from a
coupling of the inflaton to other fields $\chi$ (e.g. $\phi^2\chi^2$)
will contain a term proportional to $\phi^2T^2$ which can drive $\phi$
to the origin. Thus we should consider what happens after the Planck
era to determine whether the necessary thermal distribution can be
created from the potential energy released as the scalar fields,
initially at the Planck scale, roll towards their low energy minima.

It is important to note that the thermalisation temperature cannot be
close to the Planck scale, regardless of the amount of energy
released. To quantify this let us consider the requirements on the
inflaton field for it to be localised at the origin through its
couplings to particles in the thermal bath \cite{ggs}. On dimensional
grounds, the $2\rightarrow2$ scattering/annihilation cross-sections at
energies higher than the masses of the particles involved are expected
to decrease with increasing temperature as $\sim\alpha^2/T^2$, where
$\alpha$ is the coupling. Thus if the scattering/annihilation rate,
$\Gamma\sim\,n\langle\sigma\,v\rangle$ is to exceed the Hubble
expansion rate $H_\therm\sim(g_{\ast}T^4/10M^2)^{1/2}$ in the
radiation-dominated plasma, then we have a limit on the thermalisation
temperature $T_\therm\lesssim\alpha^2M/3g_\ast^{1/2}$, where $g_\ast$
counts the relativistic degrees of freedom (=915/4 in the minimal
supersymmetric standard model (MSSM) at high temperatures). Now a
coupling $g^2\phi^2\chi^2$ of the inflaton to another scalar field
$\chi$ will generate a confining potential at high temperatures,
$V(\phi,T)\sim\,g^2T^2\phi^2$ i.e. an effective mass for the inflaton
of $m_\therm\sim\,gT$. This will rapidly drive the inflaton field to
the origin in a time of ${\cal O}(m_\therm^{-1})$. As the universe
cools, the potential energy $\sim\Delta^4$ of the inflaton will begin
to dominate over the thermal energy at a temperature
$T_\inf\sim\Delta^2/\alpha^2M$. At this epoch the inflaton field will
be localised to a region $\delta\phi\sim\,T_\inf$ in the neighbourhood
of the origin. Thus to provide natural initial conditions for
quadratic inflation we require that:
\begin{equation}
T_\inf < T_\therm \quad {\rm i.e.} \quad \Delta \lesssim 10^{-4}M 
  \quad {\rm for} \quad \alpha \sim 1/24.  
\label{initial1}
\end{equation}
This rough estimate is consistent with the more precise explicit
calculation of the $q\bar{q}$ annihilation rate into gluons which
finds that equilibrium can only be attained below a temperature $%
T_\therm\sim3\times10^{14}$~GeV \cite{enqvist1}.\footnote{A similar
estimate of the thermalisation temperature obtains in a study where
cold particles are released at the Planck scale and allowed to scatter
to achieve equilibrium \cite{enqvist2}.} It is crucial if this
mechanism is to work that the field $\chi$ should have a mass less
than $T_\therm$ otherwise it decouples and does not contribute to the
thermal potential of the inflaton. This is a potential problem because
the inflaton vev generates a mass $g\phi$. However, as we noted above,
the term $\phi^p/M^{\prime p-4}$ does drive $\phi$ small, giving
$g\phi=gM^\prime(T/M^\prime)^{4/p}$. Thus a viable model capable of
generating the initial conditions for inflation via thermal
corrections has to satisfy the rather mild condition
\begin{equation}
gM^{\prime }(T/M^{\prime })^{4/p} < T_{\therm}.
\label{initial2}
\end{equation}

\section{Analysis of Quadratic Inflationary Models}\label{model}

In this Section we discuss the implications of quadratic inflation
using a general parameterisation of the inflationary potential. The
origin of this potential in specific theories will be discussed in
subsequent Sections. The potential can be conveniently parameterised
as a constant term driving inflation plus a quadratic term with
coefficient which may be a combination of a ``bare'' mass at the
Planck scale together with a logarithmically varying mass term
generated by radiative corrections \cite{ggs}. Thus the full potential
has the form
\begin{equation}
V(\phi) = \Delta^4\left[1 + \widetilde{b}\left(\frac{|\phi|}{M}\right)^2
 + \widetilde{c}\ln\left(\frac{|\phi|}{M}\right)
   \left(\frac{|\phi|}{M}\right)^2\right].
\label{pot1}
\end{equation}
Note that we have not included a term linear in $\phi$. As we have
stressed above such a term can be forbidden if the theory has a
symmetry under which $\phi$ transforms non-trivially and we assume
that this is the case. For example $\phi$ may be a complex field which
transforms under an Abelian symmetry (or a discrete subgroup) as
$\phi\longrightarrow{\rme}^{i\alpha}\phi$. In this case the lowest
invariant we can form is $|\phi|^2$ as in Eq.(\ref{pot1}) and $|\phi|$
is the component of $\phi$ that plays the role of the inflaton. In
what follows we shall denote the inflaton simply by $\phi$. 

In practice we are mainly interested in Eq.(\ref{pot1}) for small
$\phi\sim\phi_\H$, the field value when the density perturbations now
entering our Hubble radius exit the horizon during inflation. In this
region we can ignore, to a good approximation, the logarithmic
variation of the effective mass term and write
$V(\phi)\simeq\Delta^4(1+b\phi^2)$ where
$b=\widetilde{b}+\widetilde{c}\ln\phi_\H^2$ and $\phi$, $\Delta$ are
now expressed in Planck units: $\phi\equiv\phi/M$,
$\Delta\equiv\Delta/M$. In this case we can solve the evolution
equations analytically. After doing so we shall present numerical
solutions to the potential including the logarithmically varying terms
and compare with the analytic solution to the approximate form. Our
parameterisation of the scaled potential
$\overline{V}(\phi)=V(\phi)/\Delta^4$ is thus:
\begin{equation}
\overline{V}(\phi)
 = \left(1 - \frac{\kappa}{\Delta^q}\phi^p\right)^2 + b\phi^2 + c .  
\label{vapprox}
\end{equation}
For the case of interest, only the quadratic term is important during
inflation. However to allow us to discuss the end of inflation we have
included higher-order terms $\propto\phi^p/\Delta^q$ and above. These
terms are only relevant to the behaviour at the end of inflation;
motivation for this specific form is provided later
(Section~\ref{form}). We have also included a constant, $c$ to allow
the potential to vanish at the end of inflation. Note that a severe
fine tuning in the value of $c$ is needed to cancel the contribution
of the other terms at the true minimum of the potential. Since, to
date, there is no explanation for the observed smallness of the
cosmological constant, such a fine tuning is needed in {\em any} model
of inflation and we are not able to improve on this
situation. (However the required value of $c$ is so small that it
plays no role in determining the nature of the inflationary era.)

\subsection{Analytical solution}

We first solve Eq.(\ref{vapprox}) quite generally, without requiring
that the quadratic term dominates during inflation, although we will
be most interested later in this particular case.

\begin{itemize}

\item {\bf The end of inflation:} In the models under consideration
inflation is generated while $\phi$ rolls to larger values. The end of
inflation occurs at $\phi=\phi_{\rme}$ when the slow roll conditions
are violated. This occurs at $V^{\prime\prime}(\phi)=-\gamma$, where
$\gamma\sim1$. Thus we have
\begin{equation}
\phi_{\rme} \approx 
\left[\frac{(\gamma+2b)\Delta^q}{2\kappa p(p-1)}\right]^{1/(p-2)}.  
\label{phiend}
\end{equation}

\item {\bf Scalar density perturbations:} Solving the COBE
normalisation equation
\begin{equation}
\delta_\H^2(k) = \frac{1}{150\pi^2}\frac{V_\H^4}{\epsilon_\H}, 
\label{deltah}
\end{equation}
we find 
\begin{equation}
\phi_\H^{p-1} - \frac{b\Delta^q}{\kappa p}\phi_\H
 - \frac{\Delta^{q+2}}{2\kappa p A_\H} 
= 0,  
\label{deltadet}
\end{equation}
where $A_\H\equiv\sqrt{75}\pi\delta_\H$. This equation determines
$\Delta$ once $\phi_\H$ is determined.

\item {\bf Number of e-folds:} The number of e-folds from $\phi_\H$
to the end of inflation at $\phi_{\rme}$ is
\begin{equation}
N_\H \equiv -\int_{\phi_\H}^{\phi_\rme}\frac{V(\phi)}{V^\prime(\phi)}\rmd\phi
 \approx \int\limits_{\phi_\H}^{\phi_\rme}\rmd\phi 
 \frac{1}{-2b\phi + 2\kappa p\phi^{p-1}/\Delta^q} = 
 \frac{1}{2b(p-2)}\ln\left(\frac{1 - b\Delta^q/\kappa p\phi_\rme^{p-2}}
 {1 - b\Delta^q/\kappa p\phi_\H^{p-2}}\right).
\label{ne}
\end{equation}
Solving for $\phi_\H$ gives 
\begin{equation}
\phi_\H = \left[\frac{b\Delta^q}{\kappa p\left\{1 - \left(1 -
 \frac{2b(p-1)}{\gamma+2b}\right)\rme^{-2b(p-2)N_\H}\right\}}\right]^{1/(p-2)}
 \equiv B\Delta^{q/(p-2)}.
\label{phih}
\end{equation}
Finally substituting in Eq.(\ref{deltadet}) and simplifying we obtain the
required solution for $\Delta$: 
\begin{equation}
\Delta = \left[2\kappa pA_\H\left(B^{p-1} 
 - \frac{bB}{\kappa p}\right)\right]^{p-2/[2(p-2)-q]}. 
\label{Delta}
\end{equation}

\item {\bf Spectral Index:} This is now easily obtained to be:
\begin{eqnarray}
n_\H &\approx &1 + 2V^{\prime\prime}(\phi_\H) \nonumber \\
     &\approx &1 + 4b - 4\kappa p(p-1)B^{p-2}. 
\label{spectralindex}
\end{eqnarray}
\end{itemize}

\subsection{Numerical solution\label{numeric}}

\begin{figure}[t]
\epsfxsize14cm\epsffile{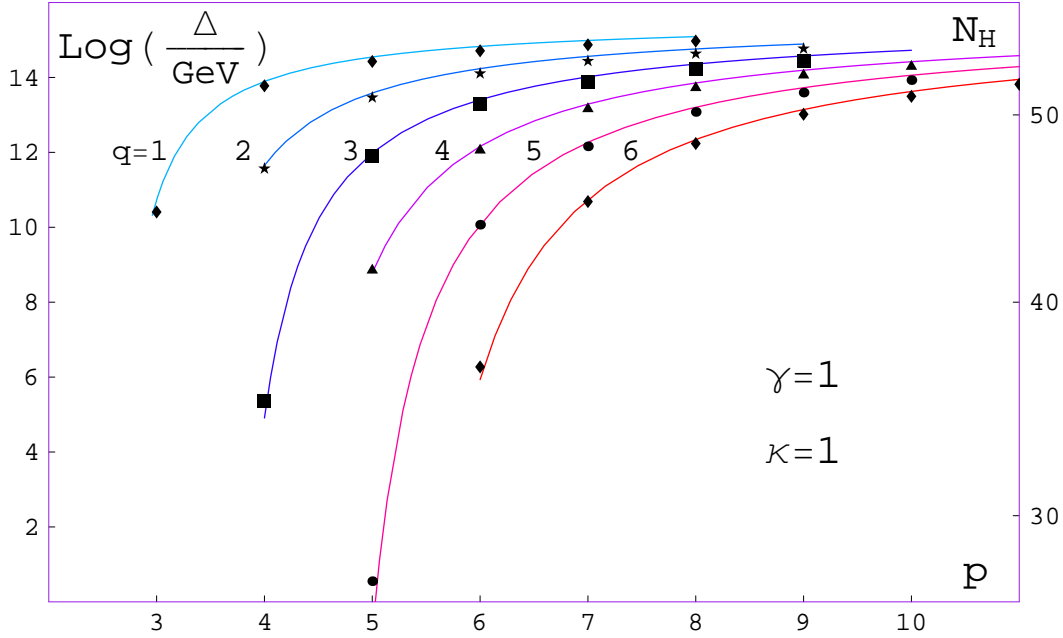}
\bigskip
\caption{The logarithm of the inflationary scale $\Delta$ as a
 function of $p$ for various values of $q$. The solid lines are the
 analytical solution (\ref{Delta}) while the symbols indicate the
 numerical results, obtained assuming $\widetilde{b}$ to be negative
 and spectral index $n_\H=0.9$. Interesting scales of inflation
 correspond in particular to the set of values
 $(p,q,\protect\kappa)=(4,2,1)$, $(4,3,1)$ and $(5,5,1)$, as shown in
 the Tables. The number of e-folds $N_\H$ indicated on the right-hand
 axis is an approximation.}
\label{fggs1}
\end{figure}

\begin{figure}[t]
\epsfxsize14cm\epsffile{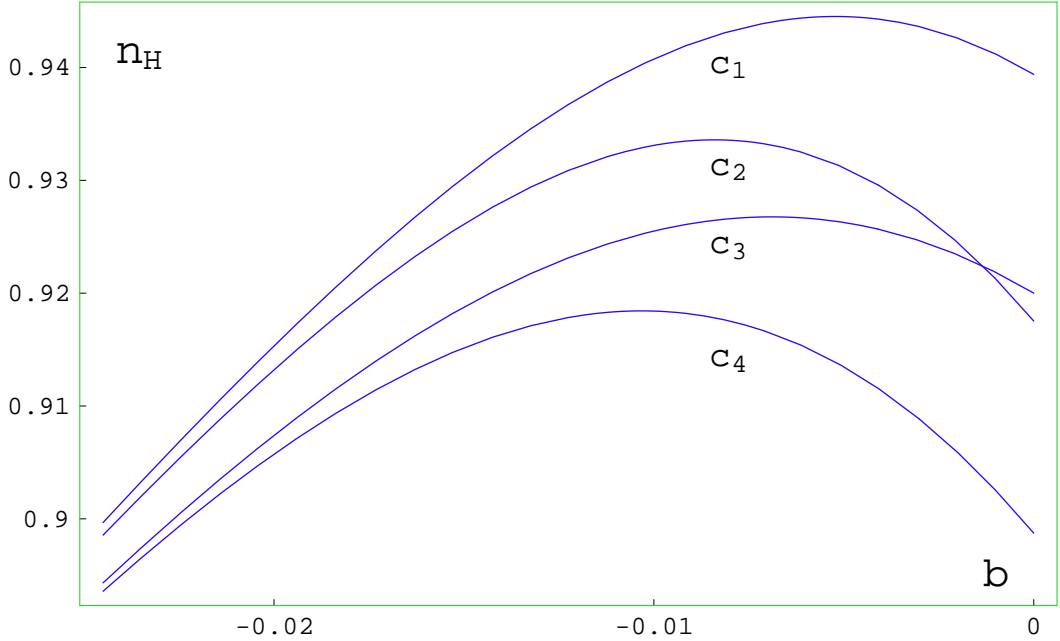}
\bigskip
\caption{The spectral index $n_\H$ (\protect\ref{spectralindex}) as a
 function of the mass parameter $b$ for the cases
 $(p,q,\kappa)=(4,2,1),(4,3,1),(5,5,1.8\times10^{-3})$ and $(5,5,1)$,
 denoted by $c_1,c_2,c_3$ and $c_4$ respectively.}
\label{fggs4}
\end{figure}

We now present numerical solutions for comparison with the analytical
results obtained above. First we show in Fig.~\ref{fggs1} the
dependence of the inflationary scale $\Delta$ on the index $p$ for
various values of $q$. The solid lines correspond to the analytical
solution (\ref{Delta}) showing how accurately these formulae reproduce
the numerical results. Fig.~\ref{fggs1} is obtained for negative
values of $\widetilde{b}$ and spectral index $n_\H=0.9$
\cite{jaffe}. We observe that interesting scales for inflation are
obtained in particular for the parameter sets $(p,q,\kappa
)=$$(4,2,1),$ $(4,3,1)$, $(5,5,1)$ and $(5,5,1.8\times 10^{-3})$, as
shown in Table~\ref{table:1}. In Fig.~\ref{fggs4} we show the
inflationary scale $\Delta$ as a function of the mass parameter $|b|$
to emphasise its insensitivity to the latter.

\section{Characteristics of Quadratic Inflation}
\label{characteristics}

Using the form for the potential (\ref{vapprox}) it is straightforward
to determine the properties of a wide variety of quadratic
inflationary models characterised by the parameters $p$, $q$ and
$\kappa$ which determine the end of inflation. In what follows we
shall usually take $\kappa=1$, its ``natural'' value and discuss a few
representative cases of the discrete possibilities specified by the
integers $p$ and $q.$ We shall also take the end of inflation to be
set by $\gamma=1$, i.e. the epoch when the curvature of the potential
becomes unity.

\subsection{The spectral index}

\begin{figure}[t]
\epsfxsize14cm\epsffile{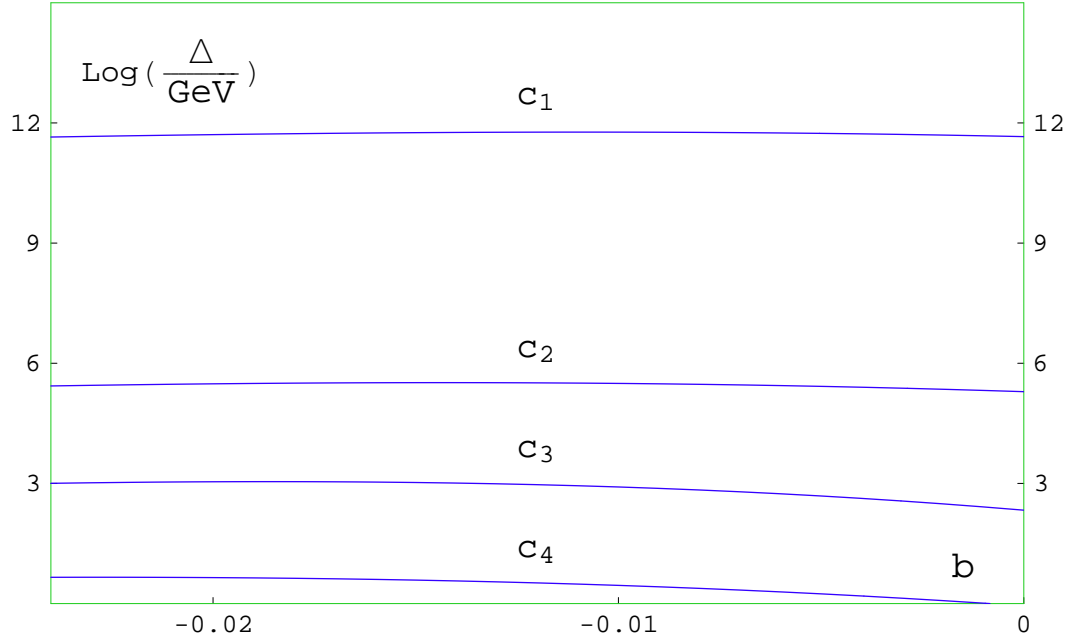}
\bigskip
\caption{The logarithm of the inflationary scale $\Delta$
 (\ref{Delta}) as a function of the mass parameter $b$ for the same
 cases as in Fig.~\protect\ref{fggs4}.}
\label{fggs5}
\end{figure}

The value of $n_\H$ is shown as a function of the mass parameter $|b|$
in Fig.~\ref{fggs5}. One can see that a characteristic property of
these models is that the spectral index is bounded as
$n_\H\lesssim0.95$ for the cases considered and that for large $|b|$
the index falls rapidly. One of the important issues is just how small
$|b|$ must be to give acceptable inflation. This gives a measure of
the severity of the $\eta$-problem and will allow us to determine
whether the candidate explanations presented in Section~\ref{eta1} do
indeed solve the problem. Using the recent observational constraints
on $n_\H$ \cite{jaffe} we can immediately obtain such a limit. In
Table~\ref{table:1} we present these limits assuming
$n_\H>0.8~(0.9)$.\footnote{The recent Boomerang and MAXIMA
observations of small angular-scale CMB anisotropy when combined with
the COBE data indicate $n_\H=0.89\pm0.06$ if the baryon to photon
ratio is fixed at the value indicated by nucleosynthesis arguments,
and $n_\H=1.01^{+0.09}_{-0.07}$ otherwise \cite{jaffe}. The COBE data
alone had previously indicated $n_\H=1.2\pm0.3$ \cite{cobe}.} We also
show the maximum value of the spectral index $n_{\H,\max}$ together
with the corresponding $|b|_\max$ for which it is reached as seen in
Fig.~\ref{fggs5}.

\subsection{Fine tuning measure}

Examination of Table~\ref{table:1} shows that the maximum value of the
effective mass squared parameter $b$ during inflation capable of
generating density perturbations in the desired range is quite
insensitive to the form of the potential at the end of inflation. As
we discuss in Section~\ref{nonfactor}, in non-factorisable higher
dimension models the value of $b$ is not constrained by the value of
the inflationary potential so there is no fine tuning implied in this
case. In supersymmetric models the ``natural'' value of $b$ is $\sim1$
so we see that a fine tuning of about 1 part in 20 is required. In our
opinion this is rather modest and can easily occur, particularly in
models in which there are several candidate inflaton fields. As
discussed in Section~\ref{eta1}, it can happen automatically in
various supergravity schemes provided the required reduction in $b$ is
relatively modest, namely $b\gtrsim\,h^2/(4\pi)^2$ where $h$ is a
coupling in the theory. We see that the required level of fine tuning
lies comfortably in this range for reasonable choices of the
coupling. We conclude therefore that it is quite likely that the
necessary conditions for quadratic inflation will be realised in
supergravity models.  Indeed in realistic models there are usually a
large number of fields associated with flat directions which are
candidates for inflatons. In this case it will be the field with the
flattest potential which will generate the last stage of inflation,
the one relevant to our observable universe.

\subsubsection{Scale of inflation}

The scale, $\Delta$, of the inflationary potential during inflation is
very sensitive to the form of the potential at the end of
inflation. However since this form is determined by the discrete
parameters $p$ and $q$, this should not be viewed as fine tuning --- a
given set will have a definite value for $\Delta$ (see
Fig.~\ref{fggs1}). We can see from Table~\ref{table:1} that a range of
$\Delta$ from 1~GeV upwards is obtained for reasonable choices of $p$
and $q$. The choice $\Delta\approx10^{11}$~GeV is particularly
interesting since one can then identify $\Delta$ with the
supersymmetry breaking scale in SUGRA models in which supersymmetry
breaking is communicated from the hidden sector to the visible sector
via gravitational strength couplings. As discussed in
Section~\ref{initial} this value is low enough for thermal effects to
set the initial value of the inflaton close to the origin, as is
required if one is to have an inflationary era. Much lower values are
possible, even down to $\sim1$~GeV and these cases are relevant to the
possibility of new large dimensions in which there is no fundamental
scale much higher than the electroweak scale.

\subsubsection{The moduli problem}

Lowering the scale of the inflationary potential also provides a
solution to the moduli problem. Moduli are scalar fields which, in the
absence of symmetry breaking triggered by non-moduli fields, have no
potential i.e.  their vevs are undetermined. They are very common in
string compactification e.g. the dilaton, the complex structure
fields, and their vevs determine the gauge and Yukawa couplings of the
theory. There are also moduli which determine the size and shape of
the compactification manifold. Because the moduli have a very flat
potential they suffer from a ``moduli'' problem due to the fact that
during inflation the minimum of the moduli potential is typically at
a different place from the minimum after inflation. Thus the moduli
fields are trapped at a false minimum during inflation and this energy
is released after inflation in the form of moduli excitations. If
these excitations become non-relativistic before they decay into
visible sector states they may release a large amount of entropy at a
late stage, unacceptably diluting both baryonic and dark matter
abundances. The reason the moduli problem may be evaded by lowering
the scale of inflation is because the distorting effect of the
inflation potential is proportional to its magnitude and so reduces as
the fourth power of the inflation scale.

To make a quantitative estimate we follow our earlier detailed
discussion \cite{ssi}. First we note that the distorting contribution
to the moduli potential is $M_{m}^2m^2$ where
$M_m^2\lesssim\,V(0)/M_\P^2$; the inequality applies because, as noted
for the inflaton, there are various possible ways the supersymmetry
breaking mass scale can be reduced. This is generated by
non-perturbative effects, such as supersymmetry breaking and may be
characterised by $\Lambda^{4}f(m/M_\P)$, where $\Lambda$ is a symmetry
breaking scale. This should now be compared to the inflaton potential
after inflation. Consider first the case $\Lambda^4<V(0)$. In this case
the moduli are lighter than the inflaton and thus decay after the
reheating epoch. As discussed earlier \cite{ssi} this leads to an
unacceptable release of entropy at late times. The problem can be
neatly avoided if $\Lambda^4>V(0)$ for in this case the moduli decay
harmlessly releasing their entropy {\em before} the inflaton decays.

What is the expectation for $\Lambda$? It has been pointed out
\cite{ssi} that for a large class of the moduli, $\Lambda$ may be
identified with the scale of symmetry breaking (triggered by
supersymmetry breaking effects) which occur below the string scale. In
models with a large intermediate scale of breaking it is quite
possible for $\Lambda^4$ to exceed $V(0)$, even in the case of linear
inflation in which $V(0)$ is very large. However at least one modulus,
the dilaton, is unaffected by such intermediate scale breaking and, as
it is unlikely that it will get a mass much larger than the
electroweak scale \cite{dilaton}, it will be lighter than the
inflaton in linear inflation models and will lead to an unacceptable
late stage of entropy release. However in quadratic inflation the
inflation scale can be very low. Indeed if $\sqrt{V(0)}/M_\P$ is less
than the electroweak scale, {\em the moduli problem may be solved even
for the dilaton}. We have seen in Section~\ref{model} that this is
indeed possible.

So far our discussion has been in the context of four-dimensional
space-time. If there are large new dimensions the moduli problem is
even more severe because the moduli associated with the size and shape
of the new dimensions cannot have mass greater than the fundamental
higher dimensional scale of the theory, which may be close to the
electroweak scale. The low inflation scale possible in
quadratic inflation is essential in these models to avoid the moduli
problem for these fields as well.

\subsubsection{Reheat temperature}

\begin{figure}[t]
\epsfxsize14cm\epsffile{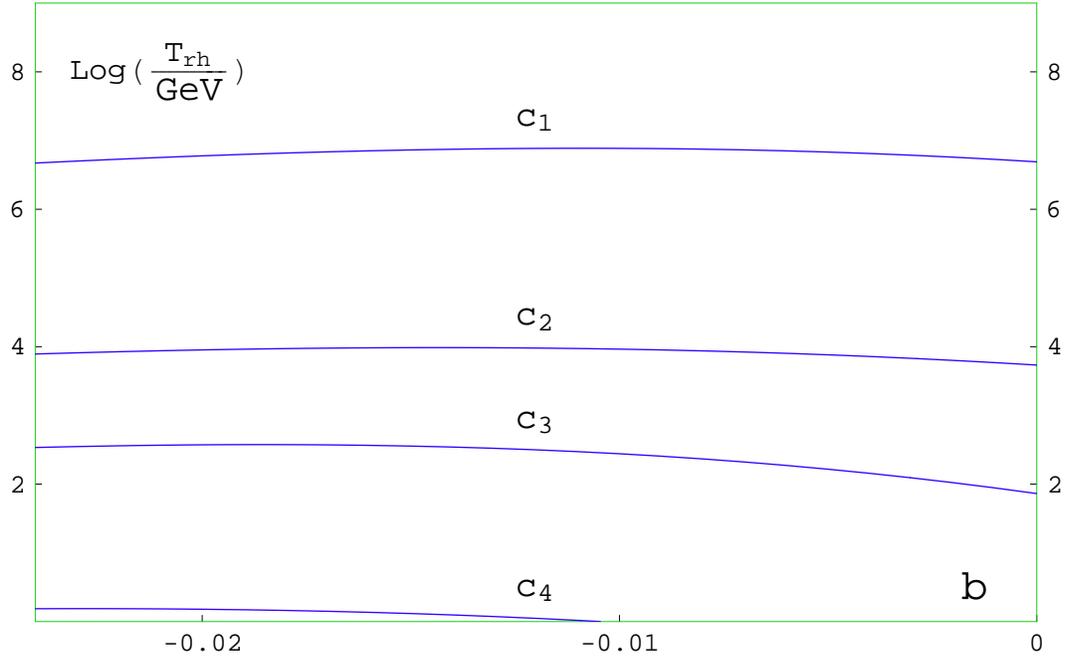}
\bigskip
\caption{The logarithm of the reheat temperature $T_\rh$
 (\protect\ref{trh}) as a function of the mass parameter $b$ for the
  same cases as in Fig.~\protect\ref{fggs4}.}
\label{fggs6}
\end{figure}

One obvious effect of lowering the scale of the inflationary potential
is a decrease of the reheat temperature. At the end of inflation the
field $\phi$ decays reheating the universe. The couplings of the
inflaton to some other bosonic $\chi$ or fermionic $\psi$ MSSM fields
occur due to terms $-\case{1}{2}g^2\phi^2\chi^2$ or
$-h\bar{\psi}\psi\phi$, respectively. These couplings induce decay
rates of the form \cite{books}
\begin{equation}
\Gamma(\phi\rightarrow\chi\chi) = \frac{g^4\phi_0^2}{8\pi m_\phi}, \qquad
 \Gamma(\phi\rightarrow\bar{\psi}\psi) = \frac{h^2m_\phi}{8\pi}\ ,
\label{decays}
\end{equation}
where $\phi_0\approx(\Delta^q/\kappa)^{1/p}$ is the value of $\phi$ at
the minimum of the potential, and $m_\phi$ is the inflaton mass given
by
\begin{equation}
m_\phi \approx \sqrt{2} p\kappa^{1/p}\Delta^{2-(q/p)}.
\label{inflamass}
\end{equation}
The decay rate is maximised when $m_{\chi,\psi}\sim\,m_\phi$ with
$\Gamma \approx\,m_\phi^3/8\pi\phi_0^2$. The reheat temperature at
the beginning of the radiation-dominated era is thus \cite{turner}
\begin{equation}
T_\rh \approx \left(\frac{90}{\pi^2g_\ast}\right)^{1/4} 
 {\min} \left(\sqrt{H(\phi_\rme)M},\sqrt{\Gamma M}\right) \approx 
 \left(\frac{30}{\pi^2g_\ast}\right)^{1/4} {\min} \left[\Delta,
 \left(\frac{3}{8\pi^2}\right)^{1/4}p^{3/2}\kappa^{5/2p}
 \Delta^{3-5q/2p}\right]\ .
\label{trh}
\end{equation}
The behaviour of $T_\rh$ as a function of the mass parameter $|b|$ is
shown in Fig.~\ref{fggs6}. Note that the observed baryon asymmetry of
the universe may in principle be generated after reheating through
anomalous electroweak $B+L$-violating processes and/or the
Affleck-Dine mechanism, even for $T_\rh$ as low as $\sim1$~GeV
\cite{baryo}.

So far we have explored the implications of quadratic inflation with a
general parameterisation of the potential, without considering its
specific origin.  We turn now to a discussion of whether such
potentials are reasonable in two attractive classes of models for
physics beyond the Standard Model.

\section{Supergravity Inflation}\label{sugrasection}

The only known {\it symmetry} capable of solving the hierarchy problem
is supersymmetry which can guarantee that a scalar mass vanishes in
the limit that supersymmetry is unbroken and ensures, through a
cancellation of bosonic and fermionic contributions, that it is not
affected by radiative corrections. However the non-vanishing potential
of Eq.(\ref{newinflation}) driving inflation {\em breaks} (global)
supersymmetry and so, even in supersymmetric models, all scalar masses
during inflation are non-zero in general. In the extreme case that the
inflaton has vanishing non-gravitational couplings, gravitational
effects will typically induce a mass of order $H_\inf\sim\Delta^2/M$
for any scalar field \cite{susybreak}, in particular the inflaton
\cite{copeland}. Nevertheless this is a big improvement over the
non-supersymmetric case, for the fine tuning problem now simply
becomes one of requiring $m_\phi=\beta \Delta^2/M$ with
$\beta\lesssim0.1$ to obtain successful inflation. In this Section we
briefly review supergravity models which can lead to this form and
discuss the structure of the resulting inflaton potential.

\subsection{The Supergravity potential}

In $N=1$ supersymmetric theories with a single SUSY generator, complex
scalar fields are the lowest components, $\phi^a$, of chiral
superfields, $\Phi^{a}$, which contain chiral fermions, $\psi^a$, as
their other components. In what follows we will take $\Phi^a$ to be
left-handed chiral superfields so that $\psi^a$ are left-handed
massless fermions.  Masses for fields will be generated by spontaneous
symmetry breakdown so that the only fundamental mass scale is the
normalised Planck scale. This is aesthetically attractive and is also
what follows if the underlying theory generating the effective
low-energy supergravity theory emerges from the superstring. The $N=1$
supergravity theory describing the interaction of the chiral
superfields is specified by the K\"{a}hler potential \cite{bailin},
\begin{equation}
G(\Phi,\Phi^\dagger) = d(\Phi,\Phi^\dagger) + \ln|f(\Phi)|^2.
\label{g}
\end{equation}
Here $d$ and $f$ (the superpotential) are two functions which need to be
specified; they must be chosen to be invariant under the symmetries of the
theory. The dimension of $d$ is 2 and that of $f$ is 3, so terms bilinear
(trilinear) in the superfields appear without any mass factors in $d$ ($f$).
The scalar potential following from Eq.(\ref{g}) is given by \cite{bailin} 
\begin{equation}
V = \rme^{d/M^2} \left[F^{A\dagger}(d_A^B)^{-1}F_B
 - 3\frac{|f|^2}{M^2}\right] + D-{\rm terms},
\label{V}
\end{equation}
where 
\begin{equation}
F_A \equiv \frac{\partial f}{\partial\Phi^A} + \left(\frac{\partial d}
 {\partial\Phi^A}\right) \frac{f}{M^2}, \qquad 
 \left(d_A^B\right)^{-1} \equiv \left(\frac{\partial^2 d} {\partial\Phi^A 
 \partial\Phi_B^\dagger}\right)^{-1}.
\end{equation}
At any point in the space of scalar fields $\Phi$ we can make a
combination of a K\"{a}hler transformation and a holomorphic field
redefinition such that $\phi^a=0$ at that point and the K\"{a}hler
potential takes the form $d=\sum_a|\Phi_a|^2+\ldots$. In this form,
the scalar kinetic terms are canonical at $\phi^a=0$ and from
Eq.(\ref{V}), neglecting $D$-terms and simplifying to the case of a
single scalar field, the scalar potential coming from the $F$-terms has
the form
\begin{equation}
V_{F} = \left( {\rme}^{|\phi|^2/M^2+\dots}\right) \left[
 |(f_{\phi}+f\phi^{\ast} + \dots)(1+\dots)|^2 - 3\frac{|f|^2}{M^2}
 \right]= V_{0} + |\phi|^2\frac{V_{0}}{M^2} + \dots, 
\label{sugra}
\end{equation}
where $V_{0}\equiv\,V_F|_{\phi=0}$.

\subsection{Supersymmetry breaking and the $\protect\eta$ problem}\label{eta1}

During inflation supersymmetry is broken by the non-zero inflaton
potential, $V$. For the case of $F$-term inflation $V_0\neq0$ and we
see from Eq.(\ref {sugra}) that the resultant breaking of
supersymmetry gives all scalar fields a contribution to their
mass-squared of $V_{0}/M^2$. If this is the only contribution to the
mass there is an obvious conflict with the slow-roll condition
(\ref{slowroll}) on $\eta$. This is the essential problem one must
solve if one is to implement inflation in a supergravity
theory. However, as stressed earlier, the problem is relatively mild
when compared to the non-supersymmetric case because the suppression
for $\eta$ need only be by a factor of 10 or so. 

There have been several proposals for dealing with this problem. One
widely explored possibility is $D$-term inflation \cite{dterm}. In
particular one may consider an anomalous $D$-term in Eq.(\ref{V}) of
the form
\begin{equation}
V_{D} = \frac{g^2}{2}(\xi-\sum_i q_i|\widetilde\phi_i|^2)^2,
\label{dterm}
\end{equation}
where $\xi$ is a constant and $\widetilde\phi_i$ are scalar fields
charged under the anomalous $U(1)$ with charge $q_i$. Such a term, for
vanishing $\widetilde\phi_i$, gives a constant term in the potential
but, unlike the case of $F$-term inflation, $V_D$ does not contribute
in leading order to the masses of uncharged scalar fields. The latter
occur in radiative order only giving a contribution to their
mass-squared equal to $\beta^2\,V_0/M^2$ where
$\beta^2\approx\,h^{\prime 2}/16\pi^2$, and typically $\beta\approx
10^{-1}$ for an effective coupling $h^\prime\sim1$ between the charged
and neutral fields. 

The other suggested ways out of the $\eta$ problem do not require
non-zero $D$-terms during inflation but rather provide reasons why the
quadratic term in Eq.(\ref{sugra}) should be anomalously small. One
possibility follows from the fact that the mass-squared term of
$V_0/M^2$ coming from Eq.(\ref{sugra}) actually applies at the Planck
scale. Since we are considering inflation for field values near the
origin the inflaton mass-squared must be run down to low scales. As
shown in Ref.\cite{kingross}, in a wide class of models in which the
gauge couplings become large at the Planck scale the low energy
supersymmetry breaking soft masses are driven much smaller at low
scales by radiative corrections. The typical effect is to reduce the
mass by a factor of $\alpha(M)/\alpha(\mu)$ where $\alpha(\mu)$ is a
gauge coupling evaluated at the scale $\mu$. While radiative
corrections can cause a significant change in the coupling, the effect
is limited and becomes smaller as the gauge coupling becomes
small. For this reason the effective mass at low scales cannot be
arbitrarily small and typically $\beta\gtrsim1/25$.

It is also possible to construct models in which the contribution to the
scalar mass exhibited in Eq.(\ref{sugra}) is cancelled by further
contributions coming from the expansion of $f$ in Eq.(\ref{V}). One
interesting example which arises in specific superstring theories was
discussed in Ref.\cite{casas}. Another occurs in supergravity models of the
`no-scale' type \cite{noscale}. In these examples, while the scalar mass is
absent in leading order, it typically arises in radiative order and so again
there is an expectation that the effective mass cannot be arbitrarily small.

To summarise, for small inflaton field values, in all these cases the
effective supergravity potential can be conveniently parameterised as a
constant term driving inflation plus a quadratic term with coefficient which
may be a combination of a ``bare'' mass at the Planck scale together with a
logarithmically varying mass term generated by radiative corrections. Thus
the full potential has just the form (\ref{pot1}) discussed earlier.

\subsection{Terminating slow-roll evolution}\label{end}

So far we have concentrated on the form of the potential which arises
due to supersymmetry breaking effects which inevitably follow from the
existence of a non-zero value for the potential during
inflation. However one may expect additional terms in the potential
which are allowed even when supersymmetry is unbroken. These terms
will determine the position of the minimum of the potential after
inflation and may also be responsible for ending inflation. 

It is this latter possibility that we shall concentrate on here
because the nature of the inflationary era is largely determined by
when inflation ends. The reason the end of inflation is likely to be
triggered by additional terms in the potential follows from a simple
dimensional argument. In the models we are considering, the inflaton
starts with field value close to the origin and rolls to larger field
values. Close to the origin the quadratic terms are likely to dominate
but as the inflaton field value increases, higher-order terms in
$\phi/M$ are likely to dominate. Such terms may occur through
higher-order terms in $d$ or $f$ in Eq.(\ref{g}) or in higher-order
terms in the expansion of Eq.(\ref{sugra}). The form of these terms is
restricted by any symmetries (discrete or continuous) of the
theory. In compactified string theories such symmetries abound,
determined by the shape of the compactification manifold. Consider
first the form of the potential following from the superpotential
$f$. In a string theory, in the absence of symmetry breaking, the
superpotential is cubic in the superfields. In the effective low
energy theory below the string and compactification scales,
higher-order terms may arise as a result of integrating out heavy
modes. The resulting form of the superpotential depends on how the
inflaton transforms under the symmetries of the theory. It is not
possible to give an exhaustive list of all possible symmetries so here
we will consider simple examples to illustrate the possibilities.

\subsubsection{Symmetries of the inflaton potential}

We start with the case where the inflaton is a singlet under all
continuous gauge symmetries but transforms under a $Z_q$ discrete
symmetry (which may be a discrete gauge symmetry and thus protected
from large gravitational corrections \cite{ibross}). The leading term
in the superpotential is then of the form $\phi^q/M^{q-3}$ where we
have taken the mass scale associated with this term to be the reduced
Planck mass, $M$. (This scale arises as a result of integrating out
heavy degrees of freedom associated with the string scale or the
compactification scale so here we are taking all such scales to be of
${\cal O}(M)$.) Following from Eq.(\ref{sugra}) we see that this term
gives rise to the leading terms of the form $|\phi|^{2q-2}/M^{2q-6}$
in the scalar potential. Note that this term is not suppressed by the
factor $\Delta^4/M^2$ associated with the quadratic
$\left|\phi\right|^2$ term because it arises even in the absence of
supersymmetry breaking. This has the important effect of reducing the
value of $\overline\phi/M$ at which the term $|\phi|^{2q-2}/M^{2q-6}$
becomes larger than the quadratic term. We see from the slow roll
conditions (\ref{slowroll}) that shortly thereafter this term will
cause inflation to end. The effect of these higher-order terms may be
enhanced still further if they arise as a result of integrating out
heavy modes {\em lighter} than the Planck scale. For example the
superpotential $f=\phi^{r}X/M^{r-2}+M_{X}X^2$ describes the
interaction of $\phi$ with a new field $X$ which has mass $M_X$. The
form of this superpotential follows from a $Z_{2r}$ symmetry under
which $\phi$ and $X$ have charge 1 and $r$ respectively. At scales
below $M_X$ the field $X$ may be integrated out to give the effective
superpotential $f=\phi^{2r}/(M^{2r-4}M_X)$. Now the scale of the
higher dimension operator is set by a combination of $M$ and $M_X$ and
for $M_X<M$ will enhance the contribution of the higher dimension
term. The new mass $M_X$ could be any of the scales in the theory,
e.g. the inflationary scale or the supersymmetry breaking scale, and
can thus be much smaller than the Planck scale.

\subsubsection{A simple parameterisation}\label{form}

With this preamble we now consider how to parameterise the structure
of the terms responsible for ending inflation. Consider first the
inflationary models driven by a non-zero $F$-term. This is conveniently
parameterised by the superpotential $f=\Delta^2Y$ which gives
$V=|F_{Y}|^2=\Delta^4$ as required. Radiative corrections then lead to
the form of Eq.(\ref{pot1}).  As discussed above we expect higher-order
corrections involving the inflaton to appear and there are many
possible forms for such corrections depending on the underlying
symmetries of the theory. Here we present a convenient
parameterisation of the potential which follows from a simple symmetry
to demonstrate how a complete inflation potential may be driven by the
symmetry structure. The starting point is the superpotential
$f=\Delta^2Y$. Such a form linear in $Y$ follows if $Y$ carries
non-zero $R$-symmetry charge $2\beta$ under an unbroken $R$-symmetry
because the full superpotential must also have charge $2\beta$ under
such a symmetry. Now let $\phi$ be a singlet under the $R$-symmetry
but have a charge under a discrete $Z_p$ symmetry. Then the most
general superpotential has the form
\begin{equation}
f = \left(\Delta^2 - \frac{\phi^p}{M^{\prime p-2}} - \frac{\phi^{2p}}
 {M^{\prime 2p-2}} - \dots \right) Y,
\end{equation}
where we have suppressed the coefficients of ${\cal O}(1)$ of each term.
This gives rise to the potential 
\begin{equation}
V = \left(\Delta^2 - \frac{\phi^p}{M^{\prime p-2}}- \frac{\phi^{2p}}
 {M^{\prime 2p-2}} - \dots \right)^2, 
\label{einf}
\end{equation}
plus terms involving $Y$ which we drop as they do not contribute to
the vacuum energy (since $Y$ does not acquire a vacuum expectation
value). For small $\phi$ the leading term has the form
$\Delta^2\phi^p/M^{\prime p-2}$ and, for the purpose of ending
inflation, this is all that matters --- the higher-order terms in
Eq.(\ref{einf}) do not play a role during inflation. Thus this form of
the superpotential gives the same inflationary phenomenology as
discussed in the previous subsection with the superpotential
$\phi^q/M^{q-3}$, provided the leading higher-order terms in $\phi$
are the same, i.e. $p=2q-2$ and
$M=M^\prime(M^{\prime}/\Delta)^{1/(q-3)}$. This illustrates the more
general point that different symmetries may lead to the same
inflationary potential. For our parameterisation we use a slightly
simplified form of Eq.(\ref{einf}) keeping only the leading $\phi^p$
term and setting $M^{\prime p-2}=\Delta^{q-2}M^{p-q}$ to take account
of the possibility discussed above that the scale associated with the
higher dimension operators may be below the Planck scale. Note $q$ is
an integer as the term $\Delta^q$ comes from heavy propagators when
integrating out massive fields. Thus we arrive at the form
\begin{equation}
V = \Delta^4\left(1 - \kappa\frac{\phi^p}{\Delta^{q}M^{p-q}}\right)^2 
 + \Delta^4\left[\widetilde{b}\left(\frac{|\phi|}{M}\right)^2
 + \widetilde{c}\ln\left(\frac{|\phi|}{M}\right)\left(\frac{|\phi|}{M}
 \right)^2\right], 
\label{param}
\end{equation}
where we have added the supersymmetry breaking terms of
Eq.(\ref{pot1}). As discussd earlier, this form of the inflationary
potential allows for a variety of inflationary scales $\Delta $
depending on the choice of $p$ and $q$ and our results for the
potential (\ref{vapprox}) in Table~\ref{table:1} should apply with the
interpretation $b=\widetilde{b}+\widetilde{c}\ln\phi_\H^2$. In
Table~\ref{table:2} we show the results of a numerical integration of
the potential (\ref{param}), demonstrating that this is indeed the
case. To conclude this discussion we show in detail how the symmetries
lead to the higher-order term $\phi^p/M^{\prime
p-2}=\Delta^{q-2}M^{p-q}$ in the superpotential for three
representative cases, identifying the origin of the mass scale
$M^\prime$ and hence the source of the $\Delta^q$ factor.

\begin{figure}[t]
\epsfxsize14cm\epsffile{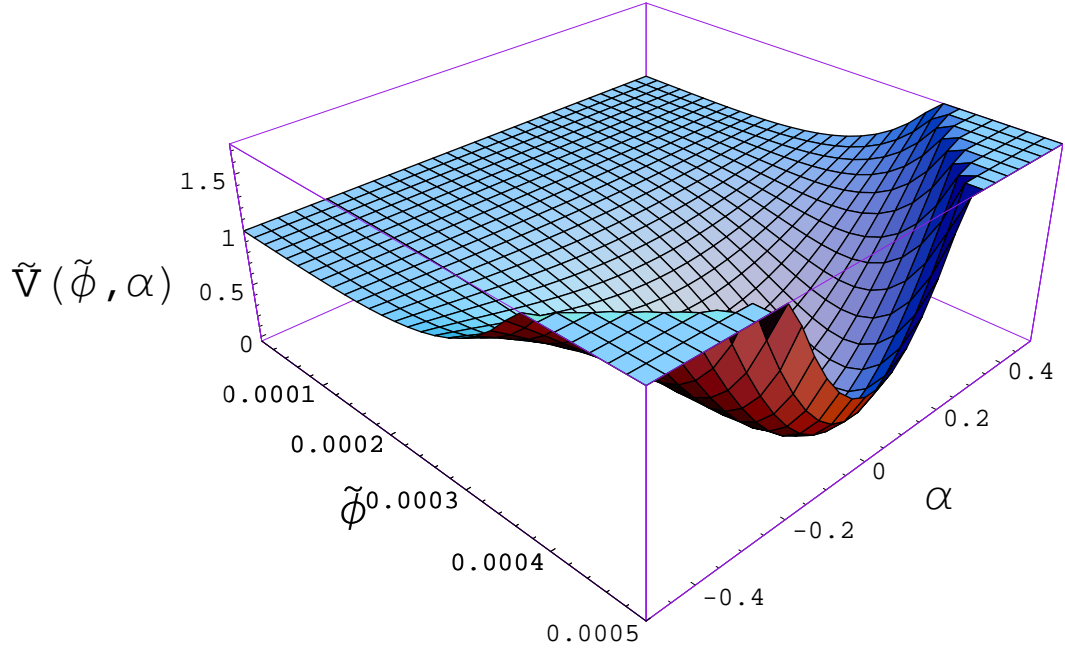}
\bigskip
\caption{The full supergravity potential (\ref{V}) (in units of
 $V_0/\Delta^4$) as a function of $\phi$ and its phase $\alpha$ for
 the case $(p,q,\kappa)=(4,2,1)$, corresponding to an inflationary
 scale of $\Delta\sim5\times10^{11}$~GeV.}
\label{fggs2}
\end{figure}

\begin{figure}[t]
\epsfxsize14cm\epsffile{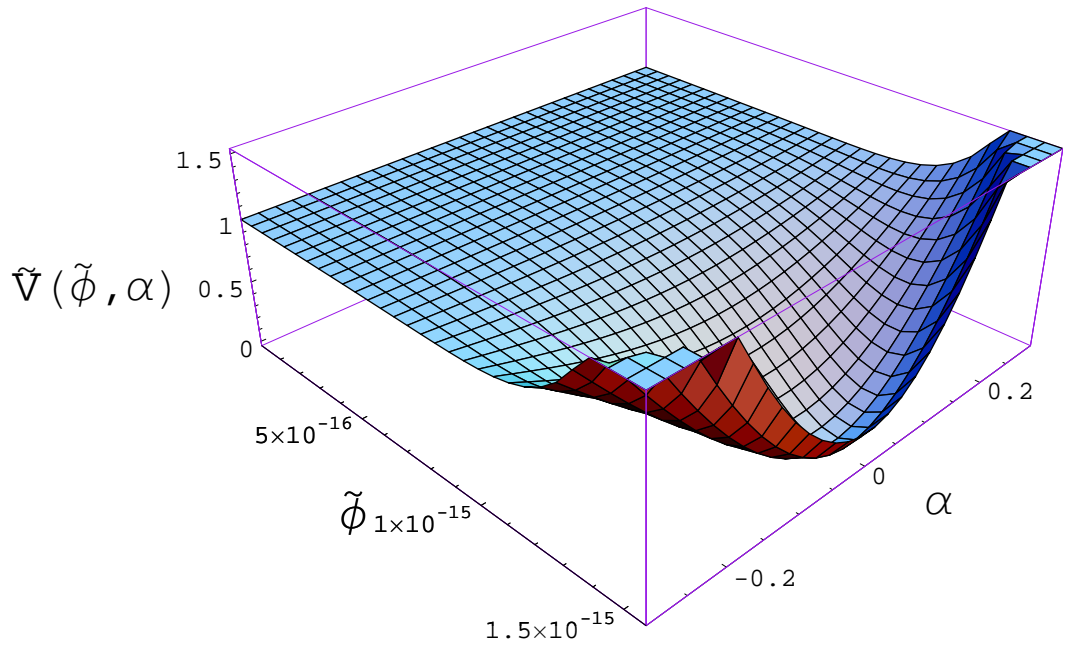}
\bigskip
\caption{Similar to Fig.~\protect\ref{fggs2}, but for the case
 $(p,q,\kappa)=(5,5,1)$ corresponding to an inflationary scale of
 $\Delta\sim1$~GeV.}
\label{fggs3}
\end{figure}

\begin{itemize}

\item {\bf Case 1:} $p=4$, $q=2$, $\Delta\approx10^{11}$~GeV

This simple case follows immediately from a $Z_4$ symmetry under which
$\phi$ has unit charge and $Y$ is neutral. Taken together with the
$R$-symmetry the leading term allowed is $Y\phi^4/M^2$ where $M$ is
the fundamental mass scale of the theory which we take to be the
Planck scale.  Combined with the supersymmetry breaking term
$Y\Delta^2$ we immediately obtain Eq.(\ref{param}) with $p=4$,
$q=2$. As discussed above, this potential gives acceptable density
perturbations provided $\Delta\approx10^{11}$~GeV, i.e. it can be
identified with the supersymmetry breaking scale. Fig.~\ref{fggs2}
shows the full supergravity inflaton potential (\ref{V}) as a function
of $\phi$ and its phase $\alpha$ for $(p,q,\kappa)=(4,2,1)$.

\item {\bf Case 2:} $p=4$, $q=3$, $\Delta\approx10^5$~GeV

In this case we choose the superpotential of the form 
\begin{equation}
f = Y\phi X + M_{X}X\overline{X} + \overline{X}\phi^{3}/M. 
\label{case2}
\end{equation}
This form arises if there is a $U(1)$ (or $Z_n$) symmetry under which the
fields $Y,X,\overline{X},\phi$ have charge -4, 3, -3 and 1 respectively.
Integrating out the massive $X,\overline{X}$ fields gives 
\begin{equation}
f = Y\phi^4/(MM_X).
\end{equation}
If $M_X=\Delta$ (as is expected if $X,\overline{X}$ belong to the
supersymmetry breaking sector driven by a gaugino condensate in which
the (confining) interactions become strong at the scale $\Delta$) we
have the desired term in the superpotential which, when combined with
the supersymmetry breaking term $Y\Delta^2$ (the SUSY breaking
condensate, $\Delta^2$, also breaks the $U(1)$ (or $Z_n$) symmetry),
yields the form given in Eq.(\ref{param}).

\item {\bf Case 3:} $p=5$, $q=5$, $\Delta\approx1$~GeV

This is readily achieved along similar lines to the previous case via the
superpotential 
\begin{equation}
f = Y\phi^2 Z + \bar{Z}X^2 + \bar{X}\phi^2 + M_{Z}Z\bar{Z}+M_{X}X\overline{X}
\end{equation}
This form arises if there is a $U(1)$ (or $Z_n$) symmetry under which
the fields $Y,Z,\bar{Z},X,\overline{X},\phi$ have charge -10, 8, -8,
4, -4 and 2 respectively. Integrating out the massive $X,\overline{X}$
fields gives
\begin{equation}
f = Y\phi^{5}/M_X^2M_{Z}
\end{equation}
With $M_X$=$M_Z$=$\Delta$ and the supersymmetry breaking term
$Y\Delta^2$ we arrive at a superpotential giving Eq.(\ref{param}) with
$p=5$, $q=5$.  Fig.~\ref{fggs3}, similar to Fig.~\ref{fggs2}, but now
for $(p,q,\kappa )=$$(5,5,1)$, shows how lower scales of inflation
require lower values of $\phi$; in particular we see how the minimum
$\phi_0$ shifts to smaller values.

\end{itemize}

\subsubsection{$D$-term inflation}

So far we have discussed the form of the inflaton potential
responsible for ending inflation that may arise as a result of
$F$-term inflation. However it is possible to obtain similar forms
using $D$-term inflation and, as noted in Section~\ref{eta1}, this may
have the advantage of eliminating the $\eta$ problem. At first sight
it seems impossible to obtain $D$-term inflation with a low scale for
the inflationary potential because the usual assumption is that the
non-zero value of the $D$-term is due to an anomalous $U(1)$ as in
Eq.(\ref{dterm}) and its scale is in turn related to the string scale.
However there are two possible ways that a low $D$-term scale may
arise. The first possibility is that the string scale itself is
low. With the realisation that the four-dimensional Planck scale may
not be a fundamental quantity has come the construction of string
models with low string scales (even as low as $\sim1$~TeV) plus large
new dimensions. In these theories the weakness of the gravitational
interactions is due to the gravitational flux spreading out in new
large extra space dimensions or through the appearance of a warp
factor in the 4-D metric dependent on the additional dimensions.
Either way the string scale and hence the associated anomalous
$D$-term is reduced. The second possibility applies even in the
original string formulations with small extra dimensions. The fields
in the hidden supersymmetry breaking sector all feel the strong
confining force responsible for gaugino condensation. Such fields may
be driven to acquire vevs of the order of the scale $\Delta$ at which
the force becomes strong. These vevs may give rise to non-zero
$D$-terms capable of generating inflation at the scale $\Delta$. Given
these possibilities what is the form of the potential and does it have
similar properties to that of Eq.(\ref{param})? In both of the cases
just discussed the general form of the $D$-term is given by
\begin{equation}
V_{D} = \left||\Delta|^2-X^{\dagger}X 
        + \overline{X}^\dagger\overline{X}\right|^2, 
\label{vd}
\end{equation}
where we have included fields $X$ and $\overline{X}$ carrying +1 and
-1 charge respectively under the $U(1)$ gauge symmetry. The end of
$D$-term inflation must be driven in a different manner from that
discussed above, namely through hybrid inflation. During inflation the
$X$ field is prevented from acquiring a vev because it has a mass,
$M_X$, which we take to come from the supersymmetry breaking sector
and hence to be of ${\cal O}(\Delta)$. The end of inflation
corresponds to the point at which the change in the vev of the
inflaton, $\phi$ , which is a singlet under the gauge symmetry
associated with the non-zero $D$-term, must alter the potential of $X$
in such a way that it can acquire a vev cancelling the $D$-term. The
most general form of the superpotential is (suppressing Yukawa
couplings)
\begin{equation}
f = M_{X}X\overline{X} + \phi X\overline{X} + {\rm higher-order\ terms} 
\label{fd}
\end{equation}
The potential following from this includes the leading terms 
\begin{equation}
V_{F} = \left|M_X+\phi\right|^2(\left|X\right|^2 
 + \left|\overline{X}\right|^2)
 +\left|X\overline{X}\right|^2.
\end{equation}
For $\phi\approx0$ the $X$ field has mass $M_X$ and for
$M_X\gtrsim\Delta$ the potential $V_D+V_F$ will constrain $X$ to have
zero vev. Once the vev of $\phi$ becomes of ${\cal O}(M_X)$ however a
cancellation of the $X$ mass term is possible and the $X$ field will
rapidly evolve so as to minimise $V_D$. Since the $X$ mass scale is
$\gtrsim\Delta$ this happens within a Hubble time and so inflation is
effectively ended. In fact for $M_X$ of ${\cal O}(\Delta)$ this
reproduces Case~1 discussed above because there too the end of
inflation occurred at $\phi$ of ${\cal O}(\Delta)$. Variations on this
hybrid theme can readily generate the other cases too. Replacing the
$\phi\overline{X}$ term in Eq.(\ref{fd}) by the term
$\phi^{s}X\overline{X}/\Delta^{t}$ (such a term can be obtained by
integrating out fields in an analogous way to that discussed above)
one finds inflation ends for
$\phi=\left(\Delta^{t}M_X\right)^{1/s}$. For the potential of the form
given in Eq.(\ref{param}) the end of inflation occurs at
$\phi\approx\Delta^{q/(p-2)}$ and so for $M_X$ of ${\cal O}(\Delta)$
and $q/(p-2)=(t+1)/s$, the end of inflation will be the same in the
$D$-term case.

\section{4-D inflation in higher dimension theories}\label{dim}

As mentioned above there has recently been much interest in a solution
to the hierarchy problem involving $\delta$ new large dimensions in
which gravity (described by closed strings in a string theory)
propagate in the $4+\delta$ dimensions while the matter states of our
world, quarks, leptons and the gauge bosons responsible for the
strong, weak and electromagnetic interactions, (described by open
string states whose ends are confined to D-branes) live in just the
normal four-dimensional space (the D=3 case). The description of
inflation in more than 4 dimensions may be done either by using the
full higher dimensional description or by using an effective 4-D
description in which the effects of the higher dimension appear as
towers of Kaluza Klein states. Which description is more appropriate
depends on the energy scale of interest. Here we use the
four-dimensional description because it is very likely that during
inflation the effective temperature drops below the compactification
scale at which the extra dimensions are frozen. Even if inflation
starts at a scale above the compactification scale for the additional
dimensions, during inflation the universe cools to the Hawking
temperature $\sim\sqrt{V(0)}/M$, and, if this is below the
compactification scale, the theory will be effectively
four-dimensional during inflation. If this is not the case we expect
problems because the phase transition corresponding to
compactification will occur {\em after} inflation. While this is not
necessarily ``no-go'', it is rather disfavoured since the moduli
setting the scale of the new dimensions propagate in the additional
dimensions and are likely to produce an unacceptable amount of entropy
in the bulk.

In the effective 4-D description, it is necessary to discuss how the
presence of additional large dimensions may change the form of the
slow-roll equations.

\subsection{Factorisable metric}

In the original realisation of this idea \cite{largedim} one assumes a
factorisable metric of the form 
\begin{equation}
\rmd s^2 = -\rmd x^0\rmd x_0 + \rmd x_i\rmd x^i + \rmd x_\alpha\rmd x^\alpha,
 \quad i=1,2,3, \quad \alpha=4,\ldots,4+\delta\ ,
\end{equation}
and finds that, due to the gravitational flux leaking out into the
extra dimensions, the gravitational coupling to matter in 4-D is much weaker
than in the higher dimensional space, with the 4-D
Planck mass given by 
\begin{equation}
M_{\P,4}^2 = M_{\P,4+\delta }^{\delta +2}\;R^{\delta} .
\label{factor}
\end{equation}
Thus provided $M_{\P,4+\delta}\;R\gg1$, one can explain why the
four-dimensional Planck mass is large while the fundamental mass scale
$M_{\P,4+\delta}$ remains small. In the extreme one may take
$M_{\P,4+\delta}$ of ${\cal O}(1)$~TeV), i.e. small enough to
eliminate the hierarchy problem which occurs when one introduces
scales much larger than the electroweak scale. In such theories there
is no mass scale higher than $M_{\P,4+\delta}$. Moreover the reheat
temperature must be extremely low \cite{largedim} if the bulk is not
to contain so much energy that it distorts the expansion rate of the
universe in an unacceptable way.  Given this it is clear that the only
viable inflationary model is one which does not require a high scale
for the inflaton potential, pointing at quadratic inflation as the
obvious candidate. Moreover it is {\em necessary} for the theory still
to be supersymmetric in order to explain why the inflaton mass is only
of order the Hubble expansion parameter during inflation (i.e. the
inflation sector supersymmetry breaking scale) --- without
supersymmetry one expects it would be driven to the scale
$M_{\P,4+\delta}$. Thus the discussion about the form of the inflation
potential in supersymmetric theories applies to factorisable metric
compactifications as well. The important difference is that in
Eq.(\ref{param}) $M$ should be identified with $M_{\P,4+\delta}$ and
not with the 4-D Planck mass. At first sight this does not seem to be
the correct prescription because the inflaton, a brane state, has only
gravitational strength couplings $\propto M_{\P,4}^{-n}$ to the new
Kaluza Klein states responsible for the additional dimensions in our
4-D description. However after summing over all the Kaluza Klein
states these couplings do induce higher-order corrections
$\propto\phi^{n+4}/M_{\P,4+\delta}^{n}$.

\subsection{Non-factorisable metric}\label{nonfactor}

Recently there has been renewed interest in the case where the metric
is not factorisable. For the case of a single extra dimension the line
element has the form
\begin{equation}
\rmd s^2 = \rme^{-\rho(r)}(-\rmd x^0\rmd x_0 + \rmd x_i\rmd x^i) +
 \rmd x_4\rmd x^4, \quad i=1,2,3, 
\label{nonfact}
\end{equation}
where $x_4=r\theta $, $0<$ $\theta \leq 2\pi $. The origin of the
hierarchy between the electroweak breaking scale and the Planck scale
is now rather different than that suggested for the factorisable
metric case. An explicit example was recently provided by Randall and
Sundrum \cite{RS1}. They considered the case of two parallel 3-branes
sitting on the fixed points of an $S^1/Z_2$ orbifold. The 5-D
spacetime is essentially a slice of five-dimensional Ant-DeSitter
space-time and the tensions of the two 3-branes are chosen so that the
4-D spacetime appears flat. This last requirement forces one of the
two branes to have negative tension. The solution to Einstein's
equations now has the form of Eq.(\ref{nonfact}) with $\rho(r)=kR$
where $k$ is the five-dimensional curvature. The exponential ``warp''
factor, $\varpi=\rme^{-kR}$, in the metric then generates a hierarchy
of the mass scales between the two branes, although all fundamental
mass scales are of order $M_{\P,4}$. The graviton is localised to the
positive tension brane while matter exists on the negative tension
brane with particle masses of order $\varpi M_{\P,4}$. Due to the
exponential dependence of the warp factor on the size of the new
dimension, one may readily generate the desired mass hierarchy with
$\varpi\sim10^{-1}$ even though the size, $R$, of the orbifold is only
$\sim30$ times $k^{-1}$ which is of order the Planck length.

It may {\em seem} that such schemes have a great advantage in
generating slow-roll inflation by easily solving the
$\eta$-problem. This follows because these models provide a reason why
the inflaton mass should be less than the Hubble parameter during
inflation even {\em without} supersymmetry. The reason is that, due to
the universal warp factor, all masses on a given brane involve the
warp factor suppression. As we have just discussed this is why on the
visible brane all masses are of order the electroweak breaking
scale. If the inflaton lives on the visible brane (or a brane very
close to it) it too will have an electroweak scale mass which is
stable against radiative corrections from sectors on other branes. In
this scheme there should be vacuum energy, $V(0)$, (the net
contribution from the bulk and the branes) driving inflation until the
inflaton rolls to its minimum to cancel the vacuum energy. Unlike the
factorisable case there is no constraint on the fundamental
higher-dimensional scale. As a result $V(0)$ can be large, even as
large as the 4-D Planck scale. Thus, in this case, both linear and
quadratic inflation appear viable even without supersymmetry.

However, if $V(0)$ is much greater than the scale, $\Lambda$, on the
inflaton brane, the inflaton will {\em not} be able to cancel $V(0)$
fully. This is because to cancel $V(0)$ the inflaton vev must be of
order $V(0)^{1/4}$. However above the scale $\Lambda$ on the inflaton
brane the underlying higher-dimension theory (string theory?) must be
used because the effective theory has uncontrolled higher-dimension
terms of the form $\phi^n/\Lambda^{n-4}$.  Thus even in this case we
need a mechanism to ensure that the inflaton is anomalously light to
satisfy the slow-roll equations (\ref{slowroll}). The obvious
candidate is supersymmetry as already discussed.

Up to now this discussion has been at a qualitative level. To be more
explicit we consider the five-dimensional case in which the Standard
Model states and the inflaton live on a 3-brane while gravity
propagates in the full five dimensions. We start with the 5-D
description. Solution of the Einstein equations for the case that the
metric is projected on to a spatially flat Friedmann-Robertson-Walker
(FRW) model and the 4-D cosmological constant after inflation is set
to zero gives a modified form for the Hubble expansion rate
\cite{braneinf,kogan,kim}
\begin{equation}
 H^2 = \frac{8\pi}{3M_4^2}\rho \left(1+\frac{\rho}{2\lambda}\right)\ , 
\label{5d}
\end{equation}
where $\lambda$ is the brane tension, $\rho$ is the energy density and
we have used $M_4$ to denote $M_{\P,4}$. It is given by
\begin{equation}
M_4 = \sqrt{\frac{3}{4\pi}}\left( \frac{M_5^2}{\sqrt{\lambda}}\right)M_5,
\end{equation}
where $M_5=M_{\P,5}$.  In Eq.(\ref{5d}) the term quadratic in
$\rho$ corrects the standard FRW cosmology. However the effective 4-D
Lagrangian should have just the Einstein-Hilbert form and one might
expect to recover the standard cosmology. In fact this is the case,
provided the Planck constant appearing therein is {\em time
dependent}.\footnote{We are grateful to A. Karch for helpful comments
concerning this point.}  It is easy to see this during inflation when
$\rho=V(\phi)$. Expanding the potential about the point where density
perturbations leave the horizon we find
$$
H^2 = \frac{1}{3M_4^2}V(0)\left(1+\frac{V(0)}{2\lambda}\right)
 \left(1+V^\prime(0)\phi + \frac{1}{2}V^{\prime\prime}(0)\phi^2+\ldots\right),
$$
where we have dropped terms of ${\cal O}(\phi/\lambda)$. One can see
that this has just the normal 4-D form but with a modified Planck mass
\begin{equation}
M_4^{\prime 2} = M_4^2 \left(1+\frac{V(0)}{2\lambda}\right)^{-1}.
\label{runningplanck}
\end{equation}
The case of chaotic (large-field) inflation (i.e. linear inflation in
our terminology) has been explored in Ref.\cite{bassett}. In this
case, for $M_{5}\ll M_4$, the new term quadratic in $V$ in
Eq.(\ref{5d}) is large, so the modified Planck mass is reduced with
$$
M_4^{\prime 2} \simeq 10^4 M_{5}^2 .
$$
As a result, following the discussion leading to Eq.(\ref{scale}), we
see that linear inflation generates the correct magnitude of density
perturbations with the scale
\begin{equation}
V(0)^{1/4} \simeq 2\sqrt{c}M_5 ,
\label{newscale}
\end{equation}
which can be low for $M_{5}$ low. Indeed in the chaotic inflationary
model investigated \cite{bassett} the scale of inflation is
$\sim\,M_{5}/10$. This demonstrates a new mechanism for generating
low-scale inflation without requiring quadratic inflation or
supersymmetry, which is apparently in conflict with the general
arguments given above!

To explain this we note that Eq.(\ref{runningplanck}) relies on the
variation of the Planck scale between the period perturbations are
produced and today, the variation corresponding to the variation of
the horizon size.  This will not be the case for compact extra
dimensions \cite{RS1} if inflation occurs after the size of the extra
dimensions are fixed because the horizon then is {\em bigger} than the
compactification radius. As argued above, we consider this to be the
likely case if the bulk is to remain empty. In this case the Planck
mass does not differ during and after inflation and, following the
arguments above, one sees that linear inflation requires the scale of
inflation to be as large as in Eq.(\ref{scale}). As a result, if one
is to have a low inflationary scale consistent with a low fundamental
scale $M_{5}$ on the brane, it is necessary to consider quadratic
inflation. Moreover, as discussed above, it is then also necessary to
invoke supersymmetry to keep the coefficient of the quadratic term
below the Hubble rate as required by the slow-roll conditions
(\ref{slowroll}).

For the case that the additional dimensions are {\it infinite}
\cite{RS2} the result of Eq.(\ref{newscale}) is a possibility, the
variation of the Planck scale corresponding to the variation of the
horizon. The new ingredient is that there is no new compact dimension
and furthermore the Planck scale is time-dependent. However even in
this case there are reasons why one needs supersymmetric quadratic
inflation. We note that the specific chaotic inflation model of
ref.\cite{bassett} requires an inflaton mass
$m\simeq5\times10^{-5}M_{5}$ and thus there must be supersymmetry or
some other mechanism responsible to keep the inflaton mass much less
than its natural scale $M_{5}$, as noted earlier \cite{dhl}. Moreover
even this is not sufficient to make linear low-scale inflation
viable. The value of the inflaton vev at the time perturbations are
produced is $\phi^\prime\sim3\times10^2M_{5}$. Since
non-renormalisable quantum corrections occur at ${\cal O}(\phi^{\prime
n+4}/M_{5}^{n})$, such large values of the inflaton field are likely
to destroy the flatness of the potential required for
inflation. Indeed, as discussed above, for $\phi'$ above $M_5$ it is
necessary to use the underlying (string?) theory which is needed to
regulate the ${\cal O}(\phi^{\prime n+4}/M_{5}^{n})$ terms. For the
case of quadratic inflation these problems need not arise because the
scale of the inflationary potential, $V(0)$, can be much smaller than
the brane tension $\lambda $ so, c.f. Eq.(\ref{5d}), the form of the
inflationary potential at the time the observable perturbations are
produced is just the normal 4-D one analysed in
Section~\ref{quadratic}.

\section{Summary}

In this paper we have explored the possibility that there could be an
inflationary era associated with a low scale for the inflationary
potential.  The most extensively studied models correspond to the case
where the inflaton potential is {\em linear} in a Taylor expansion
about the value of the inflaton field just before the time when the
observable density perturbations are produced. In these models the
scale of inflation is restricted to lie close to the Planck scale by
the requirement that scalar density perturbations should have their
observed magnitude. On the other hand, models in which the inflaton
potential is quadratic have density perturbations which depend
sensitively on the value of the inflaton field at the end of
inflation. We have explored the implications of such models in detail
and found they can generate acceptable perturbations even if the scale
of inflation is the electroweak scale or even lower.

Low scale inflation has several attractive features. It offers a
solution to the troublesome moduli problem associated with superstring
compactification. At least in the context of supersymmetric theories,
thermal effects readily impose the required initial conditions needed
for such inflation to occur. If there are large new space dimensions
solving the hierarchy problem, a low scale of inflation is essential
as there are no large fundamental mass scales.

The construction of a viable inflationary model requires an explanation
as to why the inflaton potential should be much flatter than
dimensional arguments using the fundamental mass scale of the theory
would suggest. The most promising solution is if there is an
underlying supersymmetry which maintains the flatness of the inflaton
potential even in the presence of a large (supersymmetry breaking)
cosmological (near) constant term. Simple supersymmetric models
readily lead, through large non-renormalisable terms, to the rapid end
of inflation needed to achieve low scale inflation. We have also
considered the possibility for low scale inflation in models with
large new dimensions. Although these models offer an alternative
non-supersymmetric explanation for the hierarchy problem they still
require an additional mechanism to keep the inflaton potential flat,
again suggesting the need for supersymmetry.

Determining the nature of the underlying theory leading to inflation
is a difficult task. The forthcoming precision CMB and LSS
measurements will certainly play an important role in this and we have
seen that low scale quadratic inflationary models give a
characteristic prediction (`tilted' or `red') for the spectral index
of the scalar density perturbation. Laboratory experiments will play a
complementary role because the implication of an underlying
supersymmetry is that there should be new supersymmetric states which
are likely to be observable in the next generation of
experiments. Thus there is a good chance that the ideas investigated
in this paper may be tested within the next decade.

\acknowledgements

GG gratefully acknowledges financial support from the Royal Society
and the Mexican Academy of Sciences, which made possible a visit to
the University of Oxford; he also acknowledges support by UNAM project
PAPIIT IN110200. GGR would like to thank the Aspen Institute of
Physics, where part of this work was undertaken (supported by TMR
grant FMRX-CT96-0090) and P. Binetruy, A. Karch, R. Maartens and
L. Randall for useful discussions.

\begin{table}[tbp] 
\begin{tabular}{|c|c|c|c|c|c|c|l|l|}
$p$ & $q$ & $\kappa$ & $n_\H$ & $n_{\H,\max}$ 
& $|b|$ & $|b|_\max$ & $\Delta$ (GeV) & $\Delta_\max$ (GeV) \\ \hline
4 & 2 & 1 & 0.8 (0.9) & 0.95 & 0.05 (0.024) & 0.005 & 7(46)$\times
10^{10}$ & 50$\times 10^{10}$\\ 
4 & 3 & 1 & 0.8 (0.9) & 0.93 & 0.05 (0.024) & 0.007 & 4.6(27)$\times 10^{4}$ & 28.7$\times 10^{4}$ \\ 
5 & 5 & 1 & 0.8 (0.9) & 0.92 & 0.05 (0.024) & 0.01 & 1.3(4.2) & $2.9$ \\ 
5 & 5 & 1.5$\times 10^{-3}$ & 0.8 (0.9) & 0.94 & 0.05 (0.024) & 0.008 &
1(10)$\times 10^{2}$ & 7.6$\times 10^{2}$ \\ 
\end{tabular}
\bigskip
\caption{Characteristics of representative quadratic inflation models
 for the potential (\protect\ref{vapprox}). The values of $|b|$ and
 $\Delta$ correspond to the assumed condition $n_\H>0.8 (0.9)$.}
\label{table:1}
\end{table}

\begin{table}[tbp] 
\begin{tabular}{|c|c|c|c|c|l|}
$p$ & $q$ & $\kappa$ & $n_\H$ & $|b|$ 
& $\Delta$ (GeV) \\ \hline
4 & 2 & 1 & 0.8 (0.9) & 0.05 (0.026) & $1 (3.7)\times 10^{11}$ \\ 
4 & 3 & 1 & 0.8 (0.9) & 0.05 (0.026) & $5.4 (22.4)\times10^4$ \\ 
5 & 5 & 1 & 0.8 (0.9) & 0.05 (0.025) & 1.3 (3.6) \\ 
5 & 5 & $1.5\times10^{-3}$ & 0.9 & 0.026 & $10^3$
\end{tabular}
\bigskip
\caption{Characteristics of representative quadratic inflation models
 for the potential (\protect\ref{param}) where we have chosen
 $\widetilde{b}=25\widetilde{c}$. For comparison with
 Table~\protect\ref{table:1} we give the value of
 $b=\widetilde{b}+\widetilde{c}\ln\phi_\H^2$, the effective mass at
 the epoch when the fluctuations observed by COBE leave the horizon.}
\label{table:2}
\end{table}

\end{document}